\documentclass[prl,superscriptaddress,twocolumn,nofootinbib,showpacs,preprintnumbers,floatfix]{revtex4}

\usepackage{mathrsfs}
\usepackage{amsfonts,amssymb,amsmath}
\usepackage{graphicx,epsfig}
\usepackage{multirow}
\usepackage{verbatim}
\usepackage{slashed}

\declareslashed{}{/}{0}{-0.05}{E}
\declareslashed{}{/}{0}{-0.05}{P}
\declareslashed{}{/}{0}{-0.10}{\Widehat{P}}

\def\fsu5{$\cal{F}$-$SU(5)$}

\def\bfsu5{$\boldsymbol{\mathcal{F}}$-$\boldsymbol{SU(5)}$}

\def\m1half{$M_{1/2}$}
\def\m3half{$M_{3/2}$}
\def\m32{$M_{32}$}

\def\fbns{${\rm fb}^{-1}$}
\def\mt2{$M_{T2}$}
\def\x2{$\chi^2$}
\def\2b{$M_{T2}b$}

\def\bs0{$B_S^0 \rightarrow \mu^+ \mu^-$}

\def\bea{\begin{eqnarray}}
\def\eea{\end{eqnarray}}

\begin{document}

\title{General No-Scale Supergravity: An \fsu5 Tale}

\author{Dingli Hu}

\affiliation{George P. and Cynthia W. Mitchell Institute for Fundamental Physics and Astronomy, Texas A$\&$M University, College Station, TX 77843, USA}

\author{Tianjun Li}

\affiliation{Key Laboratory of Theoretical Physics, Institute of Theoretical Physics, 
Chinese Academy of Sciences, Beijing 100190, China}

\affiliation{ School of Physical Sciences, University of Chinese Academy of Sciences, 
No.19A Yuquan Road, Beijing 100049, China}

\author{Adam Lux}

\affiliation{Department of Physics and Engineering Physics, The University of Tulsa, Tulsa, OK 74104 USA}

\author{James A. Maxin}

\affiliation{Department of Physics and Engineering Physics, The University of Tulsa, Tulsa, OK 74104 USA}

\affiliation{Department of Chemistry and Physics, Louisiana State University, Shreveport, Louisiana 71115 USA}

\author{Dimitri V. Nanopoulos}

\affiliation{George P. and Cynthia W. Mitchell Institute for Fundamental Physics and Astronomy, Texas A$\&$M University, College Station, TX 77843, USA}

\affiliation{Astroparticle Physics Group, Houston Advanced Research Center (HARC), Mitchell Campus, Woodlands, TX 77381, USA}

\affiliation{Academy of Athens, Division of Natural Sciences, 28 Panepistimiou Avenue, Athens 10679, Greece}

%%%%%%%%%%%%%%%%%%%%%%%%%%%%%%%%%%%%%%%%%%%%%%%%%%%%%%%%%%%%%%%%%%%%%%%%%%%%

\begin{abstract}

We study the grand unification model flipped $SU(5)$ with additional vector-like particle multiplets, or ${\cal F}$-$SU(5)$ for short, in the framework of General No-Scale Supergravity. In our analysis we allow the supersymmetry (SUSY) breaking soft terms to be generically non-zero, thereby extending the phenomenologically viable parameter space beyond the highly constrained one-parameter version of ${\cal F}$-$SU(5)$. In this initial inquiry, the mSUGRA/CMSSM SUSY breaking terms are implemented. We find this easing away from the vanishing SUSY breaking terms enables a more broad mass range of vector-like particles, dubbed flippons, including flippons less than 1 TeV that could presently be observed at the LHC2, as well as a lighter gluino mass and SUSY spectrum overall. This presents heightened odds that the General No-Scale ${\cal F}$-$SU(5)$ viable parameter space can be probed at the LHC2. The phenomenology comprises both bino and higgsino dark matter, including a Higgs funnel region. Particle states emerging from the SUSY cascade decays are presented to experimentally distinguish amongst the diverse phenomenological regions.

\end{abstract}

%%%%%%%%%%%%%%%%%%%%%%%%%%%%%%%%%%%%%%%%%%%%%%%%%%%%%%%%%%%%%%%%%%%%%%%%%%%%

\pacs{11.10.Kk, 11.25.Mj, 11.25.-w, 12.60.Jv}

\preprint{ACT-02-17, MI-TH-1747}

\maketitle

%%%%%%%%%%%%%%%%%%%%%%%%%%%%%%%%%%%%%%%%%%%%%%%%%%%%%%%%%%%%%%%%%%%%%%%%%%%%

\section{Introduction}

The second phase of the Large Hadron Collider (LHC) commenced in 2015, seeking to append a discovery of supersymmetry (SUSY) to the 2012 observation of the light CP-even Higgs boson. The ATLAS experiment recorded 36.0 \fbns of data in 2016 at a 13 TeV center-of-mass energy, while the CMS experiment recorded 37.82 \fbns. Given this rapid accumulation of luminosity in 2016 and soon to reenergize in 2017, the supersymmetric model space is expected to be probed beyond a 2 TeV gluino ($\widetilde{g}$) mass. The most recently published data statistics from the 2015 LHC1 run collision data of 3.9 \fbns recorded by ATLAS and 3.81 \fbns recorded by CMS provide a lower search bound of about 1.9 TeV on the gluino mass~\cite{WAdam-ICHEP}, serving as a rather strong constraint on the SUSY model space.

The beauty of supersymmetry lies in its capacity to naturally resolve several fundamental dilemmas, such as stabilization of the electroweak scale (EW), a lightest supersymmetric particle (LSP) that is stable under R-parity serving as a natural dark matter candidate, a radiative EW scale symmetry breaking mechanism, and gauge coupling unification. SUSY thus represents a promising candidate for new physics beyond the Standard Model. The SUSY search at the LHC though has returned null results thus far, with no conclusive signals yet observed. Consequently, given an experimentally measured Higgs boson mass of $m_h = 125.1$~GeV~\cite{:2012gk,:2012gu}, a rather heavy light stop ($\widetilde{t}_1)$ mass, and hence SUSY spectrum overall, is necessary in minimalistic models such as minimal Supergravity (mSUGRA) and the Constrained Minimal Supersymmetric Standard Model (CMSSM) in order to generate the required 1-loop and 2-loop contributions to the Higgs boson mass due to the large top Yukawa coupling. Accordingly, the experimentally viable SUSY spectra of mSUGRA/CMSSM, which are quite heavy, may be beyond the reach of the LHC2.

The GUT model flipped $SU(5)$ with additional vector-like multiplets, or \fsu5 for short, has been thoroughly examined in the framework of No-Scale Supergravity (SUGRA)~\cite{Maxin:2011hy,Li:2011ab,Li:2013naa,Leggett:2014hha,Li:2016bww}. In these prior analyses, the strict No-Scale SUGRA boundary conditions $M_{1/2}$ and $M_0 = A_0 = B_{\mu} = 0$ were applied. Given these rigorous constraints at the unification scale, the vector-like particle (flippon) mass scale $M_V$, top quark mass $m_t$, and low energy ratio of Higgs vacuum expectation values (VEVs) tan$\beta$ can be expressed as a function of the sole parameter $M_{1/2}$, thus serving as a true one-parameter model. While these conditions severely constrain the model space, the resulting phenomenology uncovered is quite rich. For instance, the gluino mass scale of $M_{\widetilde{g}} \ge 1.9$~TeV currently under probe at the LHC is the precise point in the No-Scale \fsu5 model space where the light Higgs boson mass enters into its experimentally viable range of $M_{h} = 125.1 \pm 0.24$~GeV, offering a plausible explanation as to why no discovery of SUSY has yet surfaced~\cite{Li:2016bww}. Furthermore, the region of the model space presently being probed by the LHC generates a relic density $\Omega h^2$ within the very narrowly constrained 9-year WMAP and Planck measurements, as well consistency with the latest experimental results of several rare decay processes and proton decay lifetimes~\cite{Li:2016bww}. Additionally, adjustments to the one-loop gauge $\beta$-function coefficients $b_i$ induced by incorporating vector-like flippon multiplets flattens the SU(3) renormalization group equation (RGE) running ($b_3$ = 0). The effective result of a vanishing $b_3$ is a lighter gluino mass and lighter spectrum overall, accelerating the LHC reach into the viable parameter space. The net consequence of such strict No-Scale conditions though is a rather massive vector-like flippon mass of $M_V \sim 23 - 50$~TeV, well beyond the reach of the current LHC and planned future upgrades.

In an effort to search the No-Scale \fsu5 model space beyond the highly constrained strict No-Scale condition $M_0 = A_0 = B_{\mu} = 0$, we now implement in this work the $General$ No-Scale SUSY breaking terms, allowing the universal scalar mass $M_0$, trilinear $A$-term coupling $A_0$, and bilinear parameter $B_{\mu}$, which is the supersymmetry breaking soft term for the $ \mu H_d H_u $ term in the superpotential, to be generically non-zero. The $M_0$ and $A_0$ terms are allowed to freely float, the values of which are solely determined by the viability of the subsequent phenomenology. On the contrary, no constraint or analysis whatsoever is placed on the parameter $B_{\mu}$. Therefore, our applied SUSY breaking terms are $M_{1/2}$, $M_0$, and $A_0$, where we implement the mSUGRA/CMSSM SUSY breaking parameters in this initial General No-Scale study. In contrast to $SU(5)$ with mSUGRA/CMSSM SUSY breaking soft terms, we expect here that the mSUGRA/CMSSM boundary conditions implemented conjointly with \fsu5 RGE running will allow a lighter gluino mass and SUSY spectrum, while the vector-like flippon Yukawa coupling to the Higgs boson can lift the Higgs mass into its experimentally preferred range, likewise permitting a light and testable SUSY spectrum at the LHC2. Moreover, less constrained SUSY breaking parameters can generate more flexibility on the vector-like mass scale $M_V$, possibly supporting production of lighter flippon masses at the LHC2.

\section{The No-Scale \bfsu5 Model}

The gauge group for minimal flipped $SU(5)$ model~\cite{Barr:1981qv, Derendinger:1983aj, Antoniadis:1987dx} is $SU(5)\times U(1)_{X}$, which may be embedded into the $SO(10)$ model. There are only two other flipped $SU(5)$ models, which are from orbifold compactification~\cite{Kim:2006hw,Huh:2009nh}. We refer the reader to Refs.~\cite{Maxin:2011hy,Li:2011ab,Li:2013naa,Leggett:2014hha,Li:2016bww} and references therein for a detailed description of the minimal flipped $SU(5)$ model. We introduce here the $XF$, $\overline{XF}$, $Xl$, and $\overline{Xl}$ vector-like particles (flippons) of Ref.~\cite{Li:2016bww} at the TeV scale to achieve string-scale gauge coupling unification~\cite{Jiang:2006hf, Jiang:2008yf, Jiang:2009za}. In string models, the masses for vector-like particles cannot be generated at stringy tree level in general since the generic superpotential is a trilinear term. Interestingly though, vector-like particle masses can be generated via instanton effects and then are exponentially suppressed. Therefore, we can obtain vector-like particle masses around the TeV scale naturally by considering proper instanton effects. The alternative method to realize TeV-scale masses for vector-like particles is the Giudice-Masiero mechanism via high-dimensional operators in the K\"ahler potential~\cite{Giudice:1988yz}.

Supersymmetry breaking must occur near the TeV scale given that mass degeneracy of the superpartners has not been observed. Supergravity models, which are GUTs with gravity mediated supersymmetry breaking, can completely characterize the supersymmetry breaking soft terms by four universal parameters (gaugino mass $M_{1/2}$, scalar mass $M_0$, trilinear soft term $A_0$, and the low energy ratio of Higgs vacuum expectation values (VEVs) $\tan\beta$), in addition to the sign of the Higgs bilinear mass term $\mu$ of the superpotential $ \mu H_d H_u $ term.

\section{Numerical Methodology}

The mSUGRA/CMSSM high-energy boundary conditions $M_{1/2},~M_0$, and $A_0$ are applied at the $M_{\cal F}$ scale near $M_{\cal F} \simeq 5 \times 10^{17}$~GeV (as opposed to an application at the traditional GUT scale of about $10^{16}$~GeV in the MSSM), along with tan$\beta$, coupled with the vector-like flippon mass decoupling scale $M_V$. The General No-Scale \fsu5 parameter space is sampled within the limits $100 \le M_{1/2} \le 5000$~GeV, $100 \le M_0 \le 5000$~GeV, $-5000 \le A_0 \le 5000$~GeV, $2 \le {\rm tan}\beta \le 65$, and $855 \le M_V \le 100,000$~GeV. The most recent LHC constraints on vector-like $T$ and $B$ quarks~\cite{atlas-vectorlike} establish lower limits of about 855~GeV for $(XQ, ~XQ^c)$ vector-like flippons and 735~GeV for $(XD, ~XD^c)$ vector-like flippons. Therefore, we set our lower $M_V$ limit at $M_V \ge 855$~GeV given that we employ a universal vector-like flippon decoupling scale. A sufficient range of the top quark mass is allowed around the world average~\cite{CDF:2013jga}, implementing liberal upper and lower limits in our analysis of $171 \le m_t \le 175$~GeV. The WMAP 9-year~\cite{Hinshaw:2012aka} and 2015 Planck~\cite{Planck:2015xua} relic density measurements are applied, where we constrain the model to be consistent with both data sets and permit the inclusion of multi-component dark matter beyond the neutralino, imposing limits of $\Omega h^2 \le 0.1300$. Consistency with the most recent LHC gluino search is strictly implemented, imposing a hard lower limit on the gluino mass in the model space of $M_{\widetilde{g}} \ge 1.9$~TeV~\cite{WAdam-ICHEP}. A lower limit on the light stop mass of $M_{\widetilde{t}_1} \ge 900$~GeV~\cite{WAdam-ICHEP} is also imposed, though the gluino constraint just noted persists as a much stronger constraint in the \fsu5 model space.

Our theoretical calculation of the light Higgs boson mass is allowed to float around the experimental central value of $m_h = 125.1$~GeV~\cite{:2012gk,:2012gu}, where we employ the larger boundaries of $123 \le m_h \le 128$~GeV to account for at least a 2$\sigma$ experimental uncertainty in addition to a theoretical uncertainty of 1.5 GeV in our calculations. The precise value of the flippon Yukawa coupling is unknown, thus we allow the coupling to span from minimal to maximal in our light Higgs boson mass calculations. At a minimal coupling, our theoretically computed light Higgs boson mass consists of only the 1-loop and 2-loop SUSY contributions, primarily from the coupling to the light stop. This computation must return a value of $m_h \le 128$~GeV, where a SUSY only contribution to the Higgs mass at this maximum of 128~GeV implies a minimal vector-like flippon contribution. At the maximal coupling, the $(XD,~XD^c)$ flippon Yukawa coupling is fixed at $Y_{XD} = 0$ and the $(XU,~XU^c)$ flippon Yukawa coupling is set at $Y_{XU} = 1$, with the $(XD,~XD^c)$ flippon trilinear coupling $A$ term set at $A_{XD} = 0$ and the $(XU,~XU^c)$ $A$ term fixed at $A_{XU} = A_U = A_0$~\cite{Huo:2011zt,Li:2011ab}. The result of the calculation assuming a maximal coupling must give $m_h \ge 123$~GeV, as this is the maximum Higgs boson mass for any particular point in the model space. Given the intersection of these dual constraints within $123 \le m_h \le 128$~GeV on our theoretical computations of the light Higgs boson mass, for each discrete point in the parameter space we simultaneously uncover both the minimally and maximally allowed Higgs boson mass when coupled to the vector-like flippons.

\begin{table*}[htp]
  \centering
  \scriptsize
  \caption{ Sample benchmark spectra for General No-Scale \fsu5. The spectra are segregated into five characteristic models based upon LSP composition: light stau coannihilation (bino LSP) with both $M_{\widetilde{t}_1} < M_{\widetilde{g}}$ and $M_{\widetilde{g}} < M_{\widetilde{t}_1}$, Higgs Funnel $(M_{H^0} \simeq 2M_{\widetilde{\chi}_1^0})$ , Higgsino LSP, and Mixed (Higgs Funnel + Higgsino). All masses are given in GeV. Those flippon masses less than 1~TeV that could presently be produced at the LHC2 are given in boldface type. The numerical values given for $\Delta a_{\mu}$ are $\times 10^{-10}$, $Br(b \rightarrow s \gamma)$ are $\times 10^{-4}$,  $Br(B_s^0 \rightarrow \mu^+ \mu^-)$ are $\times 10^{-9}$, spin-independent cross-sections $\sigma_{SI}$ are $\times 10^{-11}$~pb, and spin-dependent cross-sections $\sigma_{SD}$ are $\times 10^{-9}$~pb. The correct light Higgs boson mass around 125 GeV can be achieved by choosing the proper Yukawa coupling between the vector-like flippons and light Higgs boson, which is smaller than 1, therefore, we do not present the lightest Higgs boson mass here. }
\label{tab:spectra}
\begin{tabular}{|c||c|c|c|c|c|c||c|c|c|c|c|c|c||c|c|c|c|c|c|c|c|c|} \hline
${\rm Model}$ & $M_{1/2}$  &  $M_0$  &  $A_0$  &  $M_V$  &  tan$\beta$  &  $m_t$  &  $M_{\tilde{\chi}^0_1}$  &  $M_{\tilde{\chi}^0_2 / \tilde{\chi}^\pm_1}$  &  $M_{\tilde{\tau}_1^\pm}$  &   $M_{\tilde{t}_1}$  &  $M_{\tilde{u}_R}$  &  $M_{\tilde{g}}$  &  $M_{H^0}$  &  $\Omega h^2$  &  $\Delta a_\mu$  &  $b \to s\gamma$  &  $B_s^0 \to \mu^{\pm}$  &  $ \sigma_{SI}$  &  $\sigma_{SD}$ \\ \hline \hline
Stau & 1467 & 100 & -1060 & {\bf 855} & 18.3 & 172.9 & 293 & 628 & 297 & 1404 & 2872 & 1882 & 2850 & 0.1130 & 1.48 & 3.49 & 3.11 & 0.4 & 1\\ \hline
Stau & 1527 & 160 & -15 & {\bf 915} & 24.3 & 173.8 & 308 & 658 & 311 & 1797 & 2975 & 1974 & 2580 & 0.1110 & 1.74 & 3.51 & 3.22 & 0.5 & 2\\ \hline
Stau & 1577 & 210 & -950 & {\bf 965} & 20.4 & 174.0 & 319 & 680 & 322 & 1576 & 3063 & 2018 & 2940 & 0.1280 & 1.39 & 3.51 & 3.18 & 0.3 & 1\\ \hline
Stau & 1537 & 698 & -990 & 10825 & 34.2 & 172.6 & 344 & 714 & 349 & 1570 & 2822 & 2014 & 2220 & 0.1190 & 2.14 & 3.40 & 3.62 & 0.7 & 3 \\ \hline
Stau & 1487 & 648 & -1040 & 50373 & 33.7 & 172.8 & 353 & 725 & 357 & 1483 & 2624 & 2015 & 2060 & 0.1150 & 2.43 & 3.38 & 3.65 & 1.0 & 4 \\ \hline
Stau & 1617 & 250 & 75 & 100000 & 28.9 & 174.1 & 396 & 806 & 397 & 1802 & 2720 & 2215 & 2130 & 0.1230 & 2.25 & 3.50 & 3.34 & 1.4 & 6 \\ \hline \hline
Stau & 1527 & 160 & 970 & {\bf 915} & 28.8 & 173.5 & 308 & 659 & 311 & 2009 & 2974 & 1984 & 2290 & 0.1140 & 2.02 & 3.51 & 3.27 & 0.8 & 4\\ \hline
Stau & 1557 & 1246 & 3955 & {\bf 945} & 45.1 & 173.2 & 317 & 676 & 320 & 2500 & 3262 & 2052 & 1480 & 0.1180 & 2.53 & 3.57 & 3.73 & 5 & 25\\ \hline
Stau & 1607 & 1296 & 4005 & {\bf 995} & 45.7 & 173.0 & 328 & 700 & 331 & 2572 & 3361 & 2114 & 1510 & 0.1170 & 2.39 & 3.57 & 3.74 & 5 & 21\\ \hline
Stau & 1587 & 748 & 3000 & 10875 & 39.2 & 174.4 & 357 & 739 & 359 & 2197 & 2915 & 2095 & 1740 & 0.1160 & 2.59 & 3.55 & 3.52 & 3 & 17\\ \hline \hline
Higgs Funnel & 2483 & 4372 & 4900 & {\bf 905} & 50.0 & 172.4 & 526 & 1021 & 2795 & 4733 & 6416 & 3332 & 1120 & 0.1130 & 0.81 & 3.75 & 3.48 & 64 & 176\\ \hline
Higgs Funnel & 2483 & 4900 & 2930 & {\bf 905} & 50.0 & 171.8 & 527 & 1049 & 3328 & 4880 & 6786 & 3349 & 1030 & 0.0962 & 0.70 & 3.76 & 3.89 & 54 & 123\\ \hline
Higgs Funnel & 2493 & 4382 & 4910 & {\bf 915} & 51.0 & 174.1 & 529 & 1109 & 2690 & 4702 & 6433 & 3340 & 1050 & 0.0990 & 0.79 & 3.78 & 4.09 & 27 & 52\\ \hline
Higgs Funnel & 2543 & 4432 & 3975 & 10865 & 51.6 & 173.4 & 595 & 1206 & 2741 & 4428 & 6241 & 3348 & 1190 & 0.1120 & 0.76 & 3.71 & 4.08 & 16 & 29\\ \hline
Higgs Funnel & 1767 & 3667 & 3889 & 53333 & 51.7 & 174.1 & 431 & 859 & 2186 & 3243 & 4726 & 2484 & 928 & 0.1107 & 1.31 & 3.76 & 3.99 & 86 & 203\\ \hline
Higgs Funnel & 1772 & 3505 & 4116 & 93383 & 51.6 & 173.3 & 441 & 879 & 2044 & 3162 & 4582 & 2515 & 945 & 0.1111 & 1.37 & 3.74 & 3.88 & 77 & 177\\ \hline \hline
Higgsino & 2473 & 4890 & 4890 & {\bf 895} & 22.3 & 171.7 & 250 & 256 & 4723 & 5010 & 6790 & 3388 & 5050 & 0.0090 & 0.24 & 3.60 & 2.98 & 48 & 1750\\ \hline
Higgsino & 2493 & 4910 & 4910 & {\bf 915} & 46.6 & 172.5 & 260 & 265 & 3575 & 4987 & 6813 & 3383 & 2210 & 0.0096 & 0.52 & 3.64 & 2.88 & 52 & 1710\\ \hline
Higgsino & 2523 & 4940 & 4940 & {\bf 945} & 45.1 & 171.9 & 233 & 239 & 3717 & 5043 & 6864 & 3417 & 2580 & 0.0097 & 0.49 & 3.62 & 2.88 & 47 & 1980\\ \hline
Higgsino & 2233 & 5000 & 4556 & 80000 & 45.0 & 171.0 & 270 & 276 & 3769 & 4322 & 6225 & 3158 & 2586 & 0.0101 & 0.55 & 3.60 & 2.84 & 55 & 1670\\ \hline \hline
Mixed & 2243 & 4677 & 5010 & 40100 & 49.9 & 171.7 & 439 & 449 & 3005 & 4217 & 6006 & 3096 & 892 & 0.0019 & 0.78 & 3.84 & 2.69 & 81 & 382\\ \hline
Mixed & 1972 & 4672 & 5005 & 66717 & 50.4 & 173.3 & 434 & 453 & 2937 & 3891 & 5718 & 2792 & 884 & 0.0008 & 0.86 & 3.86 & 2.74 & 62 & 294\\ \hline
Mixed & 2105 & 5005 & 3894 & 86717 & 50.4 & 173.3 & 477 & 496 & 3296 & 4143 & 6100 & 2985 & 980 & 0.0013 & 0.75 & 3.81 & 2.88 & 86 & 406\\ \hline \hline
\end{tabular}
\end{table*}

The viable region of the model space is constrained beyond the top quark mass, light Higgs boson mass, and relic density measurements by further application of rare decay and direct dark matter detection experimental results. The rare decay experimental constraints consist of the branching ratio of the rare b-quark decay of $Br(b \to s \gamma) = (3.43 \pm 0.21^{stat}~ ±\pm 0.24^{th} \pm 0.07^{sys}) \times 10^{-4}$~\cite{HFAG}, the branching ratio of the rare B-meson decay to a dimuon of $Br(B_s^0 \to \mu^+ \mu^-) = (2.9 \pm 0.7 \pm 0.29^{th}) \times 10^{-9}$~\cite{CMS:2014xfa}, and the 3$\sigma$ intervals around the SM value and experimental measurement of the SUSY contribution to the anomalous magnetic moment of the muon of $-17.7 \times10^{-10} \le \Delta a_{\mu} \le 43.8 \times 10^{-10}$~\cite{Aoyama:2012wk}. Regarding direct dark matter detection, the constraints applied are limits on spin-independent cross-sections for neutralino-nucleus interactions derived by the Large Underground Xenon (LUX) experiment~\cite{Akerib:2016vxi} and the PandaX-II Experiment\cite{Tan:2016zwf}, and limits on the proton spin-dependent cross-sections by the COUPP Collaboration~\cite{Behnke:2012ys} and XENON100 Collaboration~\cite{Aprile:2013doa}. 

Twenty million points in the General No-Scale \fsu5 parameter space are sampled in a random scan applying the mSUGRA/CMSSM boundary conditions at the $M_{\cal F}$ scale. The SUSY mass spectra, relic density, rare decay processes, and direct dark matter detection cross-sections are calculated with {\tt MicrOMEGAs~2.1}~\cite{Belanger:2008sj} utilizing a proprietary mpi modification of the {\tt SuSpect~2.34}~\cite{Djouadi:2002ze} codebase to run flippon and General No-Scale ${\cal F}$-$SU(5)$ enhanced RGEs, utilizing non-universal soft supersymmetry breaking parameters at the scale $M_{\cal F}$. The Particle Data Group~\cite{Olive:2016xmw} world average for the strong coupling constant is $\alpha_S (M_Z) = 0.1181 \pm 0.0011$ at 1$\sigma$, and we adopt a value in this work of $\alpha_S = 0.1172$ nearer to the lower limit. 

Results of these calculations are listed in TABLE~\ref{tab:spectra} for a set of 23 viable sample benchmark points for a given set of parameters $(M_{1/2},~M_0,~A_0,~M_V,~{\rm tan}\beta,~m_t)$. The numerical relic density figures provided in TABLE~\ref{tab:spectra} consist solely of a calculation of the SUSY lightest neutralino $\widetilde{\chi}_1^0$ abundance, thus those regions with values less than the combined WMAP9 and 2015 Planck 1$\sigma$ measurement lower bound of about $\Omega h^2 \le 0.1093$ are expected to admit alternate contributions to the total observed relic density by WMAP9 and Planck. To account for possible multi-component dark matter in these regions of low neutralino density, the spin-dependent and spin-independent cross-section calculations on the \fsu5 model space shown in TABLE~\ref{tab:spectra} have been rescaled as follows:
\bea
\sigma^\textrm{re-scaled}_{SI(SD)}=\sigma_{SI(SD)}\frac{\Omega h^2}{0.1138}
\label{eq:omega}
\eea
Each of the benchmarks models in TABLE~\ref{tab:spectra} is categorized into five distinguishing regions of the viable model space identified by LSP composition. The vector-like flippon masses for the benchmark spectra in TABLE~\ref{tab:spectra} are chosen to be representative of the entire viable model space, hence we showcase both light and heavy flippon masses, between the scan limits of $855 \le M_V \le 100,000$~GeV. Vector-like flippons lighter than 1~TeV could presently be produced at the LHC2, thus we bold those $M_V$ values in TABLE~\ref{tab:spectra}.

\begin{figure}[htp]
        \centering
        \includegraphics[width=0.5\textwidth]{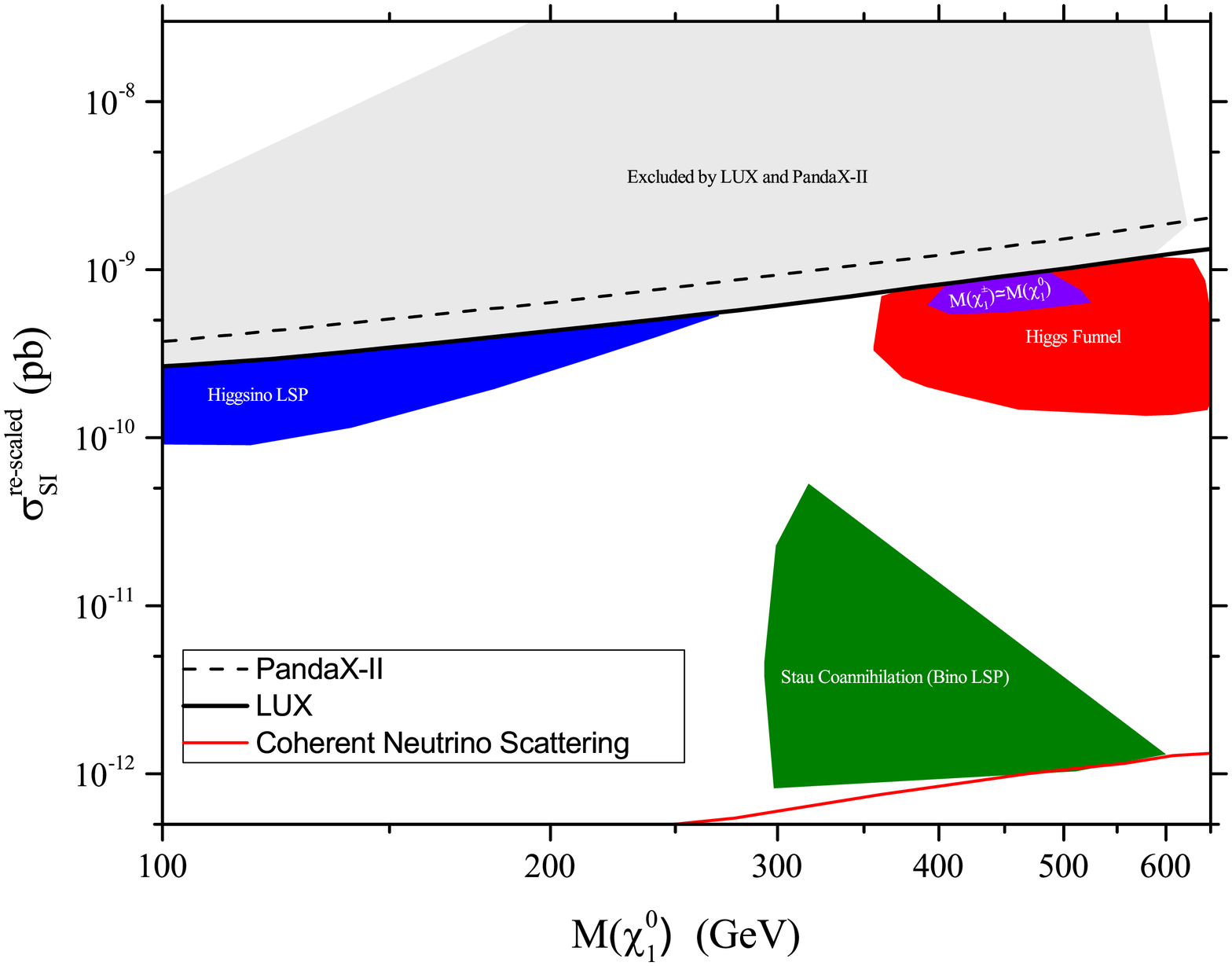}
        \caption{Illustration of the LUX and PandaX-II WIMP-nucleon spin-independent cross-section constraints applied to the General No-Scale \fsu5 viable parameter space. The null space in between the discrete Stau, Higgs Funnel, and Higgsino regions is primarily the result of application of the constraints on the gluino mass, light Higgs boson mass, and relic density.}
        \label{fig:filter}
\end{figure}

\begin{figure}[htp]
        \centering
        \includegraphics[width=0.5\textwidth]{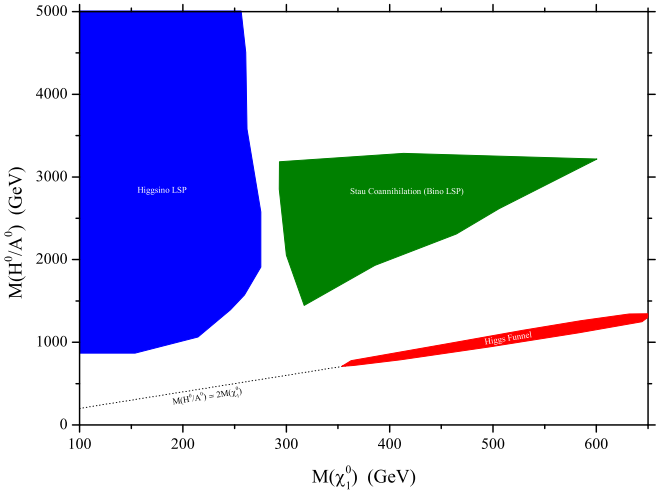}
        \caption{Depiction of $M_{H^0/A^0}$ as a function of the lightest neutralino mass $M_{\widetilde{\chi}_1^0}$. The Stau Coannihilation (bino LSP), Higgs Funnel, and Higgsino LSP regions are annotated on the plot. The linear region that defines the Higgs Funnel, namely $M_{H^0/A^0} \simeq 2 M_{\widetilde{\chi}_1^0}$, is also displayed as a dashed line.}
        \label{fig:higgs}
\end{figure}

\begin{figure}[htp]
        \centering
        \includegraphics[width=0.5\textwidth]{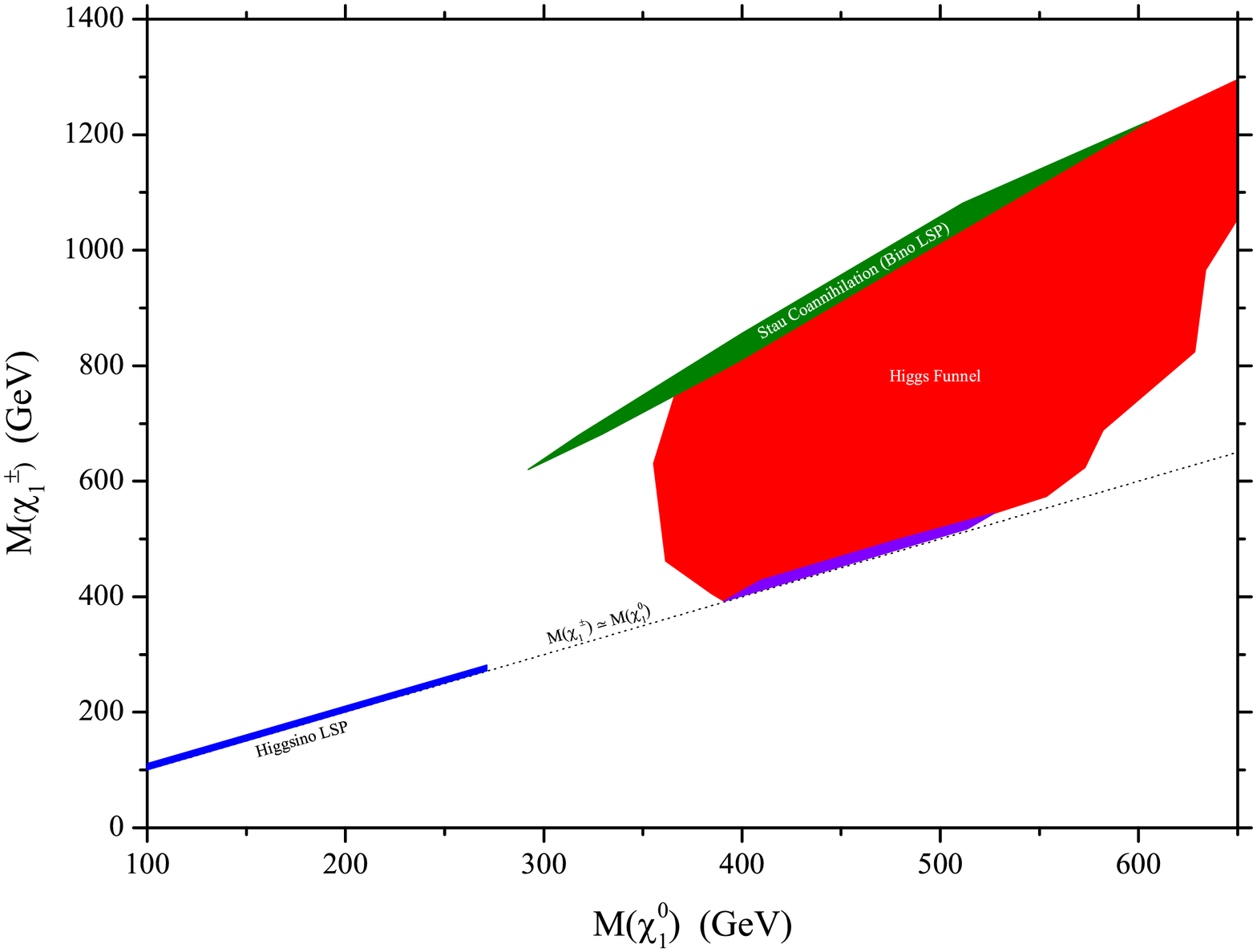}
        \caption{Depiction of $M_{\widetilde{\chi}_1^{\pm}}$ as a function of the lightest neutralino mass $M_{\widetilde{\chi}_1^0}$. The Stau Coannihilation (bino LSP), Higgs Funnel, and Higgsino LSP regions are annotated on the plot. Further highlighted here is that subspace at the lower extreme of the Higgs Funnel with $M_{\widetilde{\chi}_1^{\pm}} \simeq M_{\widetilde{\chi}_1^0}$. The linear region that defines $M_{\widetilde{\chi}_1^{\pm}} \simeq M_{\widetilde{\chi}_1^0}$ is also displayed as a dashed line.}
        \label{fig:chargino}
\end{figure}

\begin{figure}[htp]
        \centering
        \includegraphics[width=0.5\textwidth]{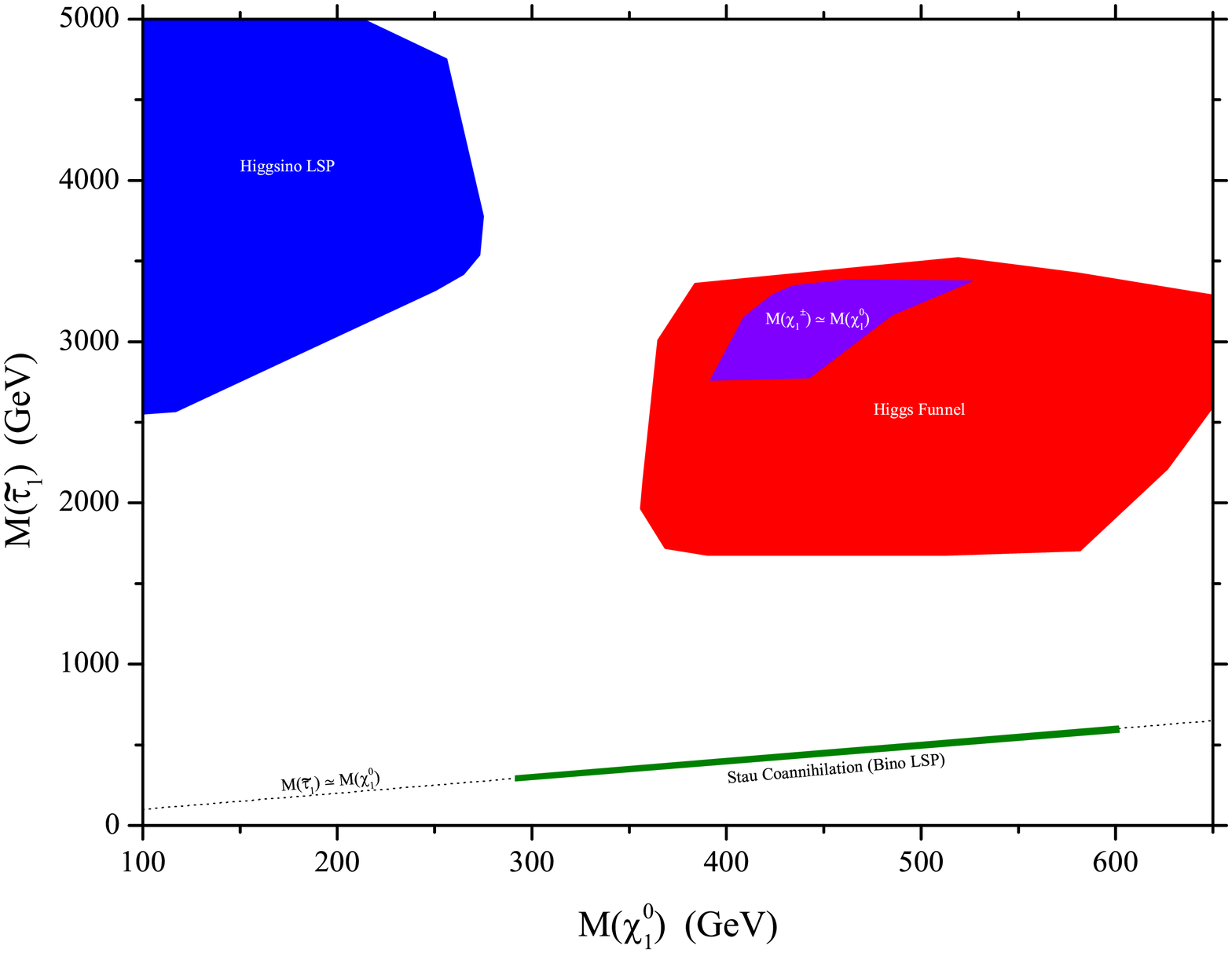}
        \caption{Depiction of $M_{\widetilde{\tau}_1^{\pm}}$ as a function of the lightest neutralino mass $M_{\widetilde{\chi}_1^0}$. The Stau Coannihilation (bino LSP), Higgs Funnel, and Higgsino LSP regions are annotated on the plot. Further highlighted here is that subspace of the Higgs Funnel with $M_{\widetilde{\chi}_1^{\pm}} \simeq M_{\widetilde{\chi}_1^0}$. The linear region that defines $M_{\widetilde{\tau}_1^{\pm}} \simeq M_{\widetilde{\chi}_1^0}$ is also displayed as a dashed line.}
        \label{fig:stau}
\end{figure}

\begin{table}[htp]
\caption{General No-Scale \fsu5 lightest supersymmetric particle (LSP) composition for the five dark matter regions studied in this work, including a comparison to the previously studied one-parameter (OPM) version of the No-Scale \fsu5 model.}
\label{tab:dm}
\begin{tabular}{|c|c|}
\hline
${\rm OPM}$ &	$100\%~{\rm bino}$ \\ \hline \hline
${\rm Stau}~(M_{\widetilde{t}_1} < M_{\widetilde{g}})$ & $100\%~{\rm bino}$ \\ \hline
${\rm Stau}~(M_{\widetilde{g}} < M_{\widetilde{t}_1})$ & $100\%~{\rm bino}$ \\ \hline
${\rm Higgs~Funnel}$ & $99\%~{\rm bino}$ \\ \hline
${\rm Higgsino}$ & $100\%~{\rm higgsino}$ \\ \hline
${\rm Mixed}$ & $98\%~{\rm higgsino}$ \\ \hline
\end{tabular}
\end{table}

\section{Phenomenological Results}

The No-Scale \fsu5 model space with the mSUGRA/CMSSM SUSY breaking terms implemented is constrained via the experimental results outlined in the prior section, with the exception of the LUX and PandaX-II spin-independent cross-sections, which we shall apply after rescaling to account for multi-component dark matter. The surviving viable parameter space consists of five distinctive regions, which we segregate based upon LSP composition. We shall show that each of these five dark matter scenarios have characteristic phenomenology and can be distinguished by means of the particle states emanating from the SUSY cascade decays. As discussed, the one-parameter version of \fsu5 generates a unique SUSY mass spectrum of $M_{\widetilde{t}_1} < M_{\widetilde{g}} < M_{\widetilde{q}}$, which does manifest again in one of the five current scenarios, though the typical mSUGRA/CMSSM SUSY spectrum of $M_{\widetilde{g}} < M_{\widetilde{t}_1} < M_{\widetilde{q}}$ is revealed also. The five regions are: (i) bino LSP with stau coannihilation and $M_{\widetilde{t}_1} < M_{\widetilde{g}} < M_{\widetilde{q}}$; (ii) bino LSP with stau coannihilation and $M_{\widetilde{g}} < M_{\widetilde{t}_1} < M_{\widetilde{q}}$; (iii) Higgs Funnel, defined as $M_{H^0} \simeq 2M_{\widetilde{\chi}_1^0}$; (iv) Higgsino LSP;  and (v) Mixed scenario, with both a Higgs Funnel and Higgsino LSP. The latter three scenarios of Higgs Funnel, Higgsino, and Mixed all possess the common SUSY spectrum mass ordering of $M_{\widetilde{g}} < M_{\widetilde{t}_1} < M_{\widetilde{q}}$, and all include regions with neutralino relic densities less than the observed value and thus would support multi-component dark matter. The LSP composition of each dark matter region is annotated in TABLE~\ref{tab:dm}. Each LSP is nearly all bino (Stau, Higgs Funnel) or all higgsino (Higgsino, Mixed). 

The five disparate regions of the model space are depicted in FIGs.~\ref{fig:filter} - \ref{fig:stau}, highlighting the stau coannihilation, Higgs Funnel, and Higgsino LSP. All of these regions are mostly segregated from each other in FIGs.~\ref{fig:filter} - \ref{fig:stau}, where the null space in between the regions is primarily the result of the application of experimental constraints on the gluino mass, light Higgs boson mass, and relic density. The latest constraints on the WIMP-nucleon spin-independent cross-sections published the LUX~\cite{Akerib:2016vxi} and PandaX-II~\cite{Tan:2016zwf} experiments are applied as a function of the LSP mass to the General No-Scale \fsu5 model in FIG.~\ref{fig:filter}. In FIG.~\ref{fig:filter} the cross-sections have been rescaled in accordance with Eq.~(\ref{eq:omega}) for those points with relic densities less than the WMAP9 and Planck observations. The LUX and PandaX-II constraints are in fact strong enough to exclude a rather large swath of the model space with spin-independent cross-sections greater than about $10^{-9}$~pb for heavier LSP masses, that would otherwise satisfy all the alternate experimental constraints applied (gluino mass, light Higgs boson mass,  relic density). Further shown is the upper boundary on coherent neutrino scattering from atmospheric neutrinos and the diffuse supernova neutrino background (DSNB), which may serve as a lower limit on direct detection probes of WIMP-nucleon scattering events. Note though that the entire viable stau coannihilation region analyzed in this work safely resides just above this neutrino scattering boundary. The LSP mass as a function of the heavy neutral Higgs pseudoscalar mass ($M_{H^0/A^0}$), the light chargino mass ($M_{\widetilde{\chi}_1^{\pm}}$), and light stau mass ($M_{\widetilde{\tau}_1^{\pm}}$) are shown in FIGs.~\ref{fig:higgs} - \ref{fig:stau}, respectively. In the Higgsino LSP scenario, it is clear from FIG.~\ref{fig:chargino} that the chargino is essentially degenerate with the LSP. 

Of particular note in TABLE~\ref{tab:spectra} are the vector-like flippon mass scales $M_V$, which are allowed to be rather light, and in fact less than 1~TeV. This is in sharp contrast to the one-parameter version of No-Scale \fsu5 where the flippons bounds must be $M_V \sim 23 - 50$~TeV, as this lighter 1~TeV mass scale affords possible production of flippons at the LHC2 (those with the $M_V$ value in boldface type in TABLE~\ref{tab:spectra}). It is also significant that several of the SUSY spectra in TABLE~\ref{tab:spectra} remain testable by the LHC2, permitting possible probing of the Stau Coannihilation, Higgs Funnel and Higgsino regions within the Run 2 schedule, a circumstance not necessarily achievable by minimal models such as $SU(5)$ with mSUGRA/CMSSM SUSY breaking soft terms that also support stau coannihilation along with a Higgs Funnel and higgsino LSP.

Testing the General No-Scale \fsu5 model requires identifying observable signatures associated with each of the five dark matter regions. The leading cascade decay channels are highlighted in TABLE~\ref{tab:br} for all regions. Note that there is only a negligible difference between the Higgsino and Mixed models with respect to the gluino branching ratios, hence we group them together in TABLE~\ref{tab:br}. Clearly the decay options proliferate for the gluino when it is lighter than the light stop, and therefore do not provide a dominant signature with which to identify that region of the parameter space. As such, the four models with $M_{\widetilde{g}} < M_{\widetilde{t}_1}$ show no channel with a branching ratio greater than about 30\%, and thus do not possess a dominant decay mode. On the contrary, the gluino in the one-parameter version of No-Scale \fsu5 will decay to a light stop and hence $t \bar{t}$ 100\% of the time~\cite{Li:2013naa} and the stau coannihilation region of General No-Scale SUGRA with $M_{\widetilde{t}_1} < M_{\widetilde{g}}$ also shows a reasonably large branching ratio of about 62\% to a $t \bar{t}$, thus these provide a much stronger singular decay channel. The fact that the $t \bar{t}$ production does vary considerably between the five different regions does though present an opportunity to utilize the $t \bar{t}$ channel as a tool with which to discriminate between all the regions, and furthermore, differentiate the one-parameter version of No-Scale \fsu5~\cite{Li:2016bww} from General No-Scale \fsu5. The $\widetilde{g} \to t \bar{t}$ branching ratios are itemized in TABLE~\ref{tab:top}, displaying very evidently the divergence in this channel. Each gluino that results in a $t \bar{t}$ can produce up to six hadronic jets, with two b-jets among them, therefore manifesting as a large multijet event at the LHC, particularly with pair-produced gluinos. Large multijet events are the characteristic signature of the one-parameter version of No-Scale \fsu5~\cite{Maxin:2011hy}, and TABLE~\ref{tab:top} shows that the number of multijet events could potentially be used to identify all the models studied here.

\begin{table}
\caption{General No-Scale \fsu5 leading cascade decay channels for the five different dark matter regions studied in this work. The BR column represents the branching ratio.}
\label{tab:br}
\begin{tabular}{|c|c|c|}
\hline
${\rm Model}$&${\rm ~~BR~~}$ & ${\rm Decay~Mode}$ \\ \hline \hline
${\rm Stau}~(M_{\widetilde{t}_1} < M_{\widetilde{g}})$ & $0.62$ & $ \widetilde{g} \rightarrow   t\bar{t} + \widetilde{\chi}_1^0$ \\ \hline 
${\rm Stau}~(M_{\widetilde{t}_1} < M_{\widetilde{g}})$ & $0.11$ & $ \widetilde{g} \rightarrow  tb + \tau + \nu_\tau + \widetilde{\chi}_1^0$ \\ \hline \hline \hline
${\rm Stau}~(M_{\widetilde{g}} < M_{\widetilde{t}_1})$ & $0.31$ & $ \widetilde{g} \rightarrow   t \bar{t} + \widetilde{\chi}_1^0$ \\ \hline 
${\rm Stau}~(M_{\widetilde{g}} < M_{\widetilde{t}_1})$ & $0.19$ & $ \widetilde{g} \rightarrow  tb + \tau +  \nu_\tau + \widetilde{\chi}_1^0$ \\ \hline 
${\rm Stau}~(M_{\widetilde{g}} < M_{\widetilde{t}_1})$ & $0.19$ & $ \widetilde{g} \rightarrow   q\bar{q} + \tau + \nu_\tau + \widetilde{\chi}_1^0$ \\ \hline 
${\rm Stau}~(M_{\widetilde{g}} < M_{\widetilde{t}_1})$ & $0.15$ & $ \widetilde{g} \rightarrow  q\bar{q} + \tau^+ \tau^- + \widetilde{\chi}_1^0$ \\ \hline \hline \hline
${\rm Higgs~Funnel}$ & $0.30$ & $ \widetilde{g} \rightarrow  tb + W + \widetilde{\chi}_1^0$ \\ \hline 
${\rm Higgs~Funnel}$ & $0.11$ & $ \widetilde{g} \rightarrow  t \bar{t} + Z + \widetilde{\chi}_1^0$ \\ \hline 
${\rm Higgs~Funnel}$ & $0.09$ & $ \widetilde{g} \rightarrow  t \bar{t} + h + \widetilde{\chi}_1^0$ \\ \hline \hline \hline
${\rm Higgsino/Mixed}$ & $0.28$ & $ \widetilde{g} \rightarrow   tb + q\bar{q}+ \widetilde{\chi}_1^0$ \\ \hline 
${\rm Higgsino/Mixed}$ & $0.12$ & $ \widetilde{g} \rightarrow  t \bar{t} + \widetilde{\chi}_1^0$ \\ \hline 
\end{tabular}
\end{table}

\begin{table}[htp]
\caption{Branching ratios of $\tilde{g} \rightarrow t \bar{t}+\chi_1^0$ in General No-Scale \fsu5. The models represent the five different dark matter regions studied in this work, including a comparison to the previously studied one-parameter (OPM) version of No-Scale \fsu5. Note that the contrasting level of $t \bar{t}$ production in each of the five regions can be utilized to discriminate amongst the models.}
\label{tab:top}
\begin{tabular}{|c|c|}
\hline
${\rm Model}$ & ${\rm Br}(\widetilde{g} \rightarrow t \bar{t} + \widetilde{\chi}_1^0)$\\ \hline \hline
${\rm OPM}$ &	$1.00$ \\ \hline \hline
${\rm Stau}~(M_{\widetilde{t}_1} < M_{\widetilde{g}})$ & $0.62$ \\ \hline
${\rm Stau}~(M_{\widetilde{g}} < M_{\widetilde{t}_1})$ & $0.31$ \\ \hline
${\rm Higgsino/Mixed}$ & $0.12$ \\ \hline
${\rm Higgs~Funnel}$ & $0.03$ \\ \hline
\end{tabular}
\end{table}

While the gluino decay modes can be utilized to discriminate amongst the regions of varied dark matter scenarios, in contrast, the squark channels are reasonably consistent throughout the model space. Identifying $\widetilde{q} = (\widetilde{u}, \widetilde{d}, \widetilde{c}, \widetilde{s})$, and simply computing an approximate average between branching ratios of right-handed squarks $\widetilde{q}_R$ and left-handed squarks $\widetilde{q}_L$, we find a mean branching ratio for $\widetilde{q} \to \widetilde{g} + q$ of about 75\%. On the other hand, the light stop decay modes are more diverse given that it can be lighter or heavier than the gluino. Similar to the one-parameter version of No-Scale \fsu5, the light stop in the General No-Scale \fsu5 Stau regions will produce a top quark via $\widetilde{t}_1 \to t + \widetilde{\chi}_1^0$ 100\% of the time. This can be attributed to the light stop being lighter than the gluino, or in the case of the Stau region with $M_{\widetilde{g}} < M_{\widetilde{t}_1}$, the mass delta is rather small with the two sparticles nearly degenerate. The situation is not as clean in the remaining model space where the gluino is much lighter than the light stop, as the primary channel for the light stop in each region will be $\widetilde{t}_1 \to \widetilde{g} + t$ at 40\% (Higgsino), 36\% (Higgs Funnel), and 31\% (Mixed).

An intriguing aspect to recognize in TABLE~\ref{tab:br} regards the $t \bar{t} h$ state in the Higgs Funnel, which is the production of a light Higgs boson in tandem with a $t \bar{t}$. The $t \bar{t} h$ production cross-section via off-shell top quarks in the Standard Model is well known and is used as a direct measurement of the tree-level top Yukawa coupling. This places strong limits on supersymmetric contributions to gluon fusion processes, and interestingly, strong $t \bar{t} h$ production has been observed at the LHC~\cite{Khachatryan:2016vau}. While the branching ratio $t \bar{t} h$ in the General No-Scale \fsu5 Higgs Funnel is a mere 9\%, this does suggest possible non-negligible production of these events in the current LHC Run 2.

%%%%%%%%%%%%%%%%%%%%%%%%%%%%%%%%%%%%%%%%%%%%%%%%%%%%%%%%%%%%%%%%%%%%%%%%%%%%

\section{Acknowledgments}

The computing for this project was performed at the Tandy Supercomputing Center, using dedicated resources provided by The University of Tulsa. This research was supported in part by the Projects 11475238 and 11647601 supported by the National Natural Science Foundation of China, and by the DOE grant DE-FG02-13ER42020 (DVN). 

%%%%%%%%%%%%%%%%%%%%%%%%%%%%%%%%%%%%%%%%%%%%%%%%%%%%%%%%%%%%%%%%%%%%%%%%%%%%

\bibliography{bibliography}

\begin{thebibliography}{33}
\expandafter\ifx\csname natexlab\endcsname\relax\def\natexlab#1{#1}\fi
\expandafter\ifx\csname bibnamefont\endcsname\relax
  \def\bibnamefont#1{#1}\fi
\expandafter\ifx\csname bibfnamefont\endcsname\relax
  \def\bibfnamefont#1{#1}\fi
\expandafter\ifx\csname citenamefont\endcsname\relax
  \def\citenamefont#1{#1}\fi
\expandafter\ifx\csname url\endcsname\relax
  \def\url#1{\texttt{#1}}\fi
\expandafter\ifx\csname urlprefix\endcsname\relax\def\urlprefix{URL }\fi
\providecommand{\bibinfo}[2]{#2}
\providecommand{\eprint}[2][]{\url{#2}}

\bibitem[{\citenamefont{Adam}(2016)}]{WAdam-ICHEP}
\bibinfo{author}{\bibfnamefont{W.}~\bibnamefont{Adam}},
  {``}\bibinfo{title}{{Searches for SUSY, talk at the 38th International
  Conference on High Energy Physics}},{''} (\bibinfo{year}{2016}).

\bibitem[{\citenamefont{Aad et~al.}(2012)}]{:2012gk}
\bibinfo{author}{\bibfnamefont{G.}~\bibnamefont{Aad}} \bibnamefont{et~al.}
  (\bibinfo{collaboration}{ATLAS Collaboration}),
  {``}\bibinfo{title}{{Observation of a new particle in the search for the
  Standard Model Higgs boson with the ATLAS detector at the LHC}},{''}
  \bibinfo{journal}{Phys.Lett.} \textbf{\bibinfo{volume}{B716}},
  \bibinfo{pages}{1} (\bibinfo{year}{2012}), \eprint{1207.7214}.

\bibitem[{\citenamefont{Chatrchyan et~al.}(2012)}]{:2012gu}
\bibinfo{author}{\bibfnamefont{S.}~\bibnamefont{Chatrchyan}}
  \bibnamefont{et~al.} (\bibinfo{collaboration}{CMS Collaboration}),
  {``}\bibinfo{title}{{Observation of a new boson at a mass of 125 GeV with the
  CMS experiment at the LHC}},{''} \bibinfo{journal}{Phys.Lett.}
  \textbf{\bibinfo{volume}{B716}}, \bibinfo{pages}{30} (\bibinfo{year}{2012}),
  \eprint{1207.7235}.

\bibitem[{\citenamefont{Li et~al.}(2011)\citenamefont{Li, Maxin, Nanopoulos,
  and Walker}}]{Maxin:2011hy}
\bibinfo{author}{\bibfnamefont{T.}~\bibnamefont{Li}},
  \bibinfo{author}{\bibfnamefont{J.~A.} \bibnamefont{Maxin}},
  \bibinfo{author}{\bibfnamefont{D.~V.} \bibnamefont{Nanopoulos}},
  \bibnamefont{and} \bibinfo{author}{\bibfnamefont{J.~W.}
  \bibnamefont{Walker}}, {``}\bibinfo{title}{{The Ultrahigh jet multiplicity
  signal of stringy no-scale ${\cal F}$-$SU(5)$ at the $\sqrt{s}= 7$ TeV
  LHC}},{''} \bibinfo{journal}{Phys.Rev.} \textbf{\bibinfo{volume}{D84}},
  \bibinfo{pages}{076003} (\bibinfo{year}{2011}), \eprint{1103.4160}.

\bibitem[{\citenamefont{Li et~al.}(2012)\citenamefont{Li, Maxin, Nanopoulos,
  and Walker}}]{Li:2011ab}
\bibinfo{author}{\bibfnamefont{T.}~\bibnamefont{Li}},
  \bibinfo{author}{\bibfnamefont{J.~A.} \bibnamefont{Maxin}},
  \bibinfo{author}{\bibfnamefont{D.~V.} \bibnamefont{Nanopoulos}},
  \bibnamefont{and} \bibinfo{author}{\bibfnamefont{J.~W.}
  \bibnamefont{Walker}}, {``}\bibinfo{title}{{A Higgs Mass Shift to 125 GeV and
  A Multi-Jet Supersymmetry Signal: Miracle of the Flippons at the $\sqrt{s} =
  7$~TeV LHC}},{''} \bibinfo{journal}{Phys.Lett.}
  \textbf{\bibinfo{volume}{B710}}, \bibinfo{pages}{207} (\bibinfo{year}{2012}),
  \eprint{1112.3024}.

\bibitem[{\citenamefont{Li et~al.}(2013)\citenamefont{Li, Maxin, Nanopoulos,
  and Walker}}]{Li:2013naa}
\bibinfo{author}{\bibfnamefont{T.}~\bibnamefont{Li}},
  \bibinfo{author}{\bibfnamefont{J.~A.} \bibnamefont{Maxin}},
  \bibinfo{author}{\bibfnamefont{D.~V.} \bibnamefont{Nanopoulos}},
  \bibnamefont{and} \bibinfo{author}{\bibfnamefont{J.~W.}
  \bibnamefont{Walker}}, {``}\bibinfo{title}{{No-Scale ${\cal F}$-$SU(5)$ in
  the Light of LHC, Planck and XENON}},{''} \bibinfo{journal}{Jour.Phys.}
  \textbf{\bibinfo{volume}{G40}}, \bibinfo{pages}{115002}
  (\bibinfo{year}{2013}), \eprint{1305.1846}.

\bibitem[{\citenamefont{Leggett et~al.}(2015)\citenamefont{Leggett, Li, Maxin,
  Nanopoulos, and Walker}}]{Leggett:2014hha}
\bibinfo{author}{\bibfnamefont{T.}~\bibnamefont{Leggett}},
  \bibinfo{author}{\bibfnamefont{T.}~\bibnamefont{Li}},
  \bibinfo{author}{\bibfnamefont{J.~A.} \bibnamefont{Maxin}},
  \bibinfo{author}{\bibfnamefont{D.~V.} \bibnamefont{Nanopoulos}},
  \bibnamefont{and} \bibinfo{author}{\bibfnamefont{J.~W.}
  \bibnamefont{Walker}}, {``}\bibinfo{title}{{Confronting Electroweak
  Fine-tuning with No-Scale Supergravity}},{''} \bibinfo{journal}{Phys.Lett.}
  \textbf{\bibinfo{volume}{B740}}, \bibinfo{pages}{66} (\bibinfo{year}{2015}),
  \eprint{1408.4459}.

\bibitem[{\citenamefont{Li et~al.}(2017)\citenamefont{Li, Maxin, and
  Nanopoulos}}]{Li:2016bww}
\bibinfo{author}{\bibfnamefont{T.}~\bibnamefont{Li}},
  \bibinfo{author}{\bibfnamefont{J.~A.} \bibnamefont{Maxin}}, \bibnamefont{and}
  \bibinfo{author}{\bibfnamefont{D.~V.} \bibnamefont{Nanopoulos}},
  {``}\bibinfo{title}{{The return of the King: No-Scale ${\cal
  F}$-$SU(5)$}},{''} \bibinfo{journal}{Phys. Lett.}
  \textbf{\bibinfo{volume}{B764}}, \bibinfo{pages}{167} (\bibinfo{year}{2017}),
  \eprint{1609.06294}.

\bibitem[{\citenamefont{Barr}(1982)}]{Barr:1981qv}
\bibinfo{author}{\bibfnamefont{S.~M.} \bibnamefont{Barr}},
  {``}\bibinfo{title}{{A New Symmetry Breaking Pattern for $SO(10)$ and Proton
  Decay}},{''} \bibinfo{journal}{Phys. Lett.} \textbf{\bibinfo{volume}{B112}},
  \bibinfo{pages}{219} (\bibinfo{year}{1982}).

\bibitem[{\citenamefont{Derendinger et~al.}(1984)\citenamefont{Derendinger,
  Kim, and Nanopoulos}}]{Derendinger:1983aj}
\bibinfo{author}{\bibfnamefont{J.~P.} \bibnamefont{Derendinger}},
  \bibinfo{author}{\bibfnamefont{J.~E.} \bibnamefont{Kim}}, \bibnamefont{and}
  \bibinfo{author}{\bibfnamefont{D.~V.} \bibnamefont{Nanopoulos}},
  {``}\bibinfo{title}{{Anti-$SU(5)$}},{''} \bibinfo{journal}{Phys. Lett.}
  \textbf{\bibinfo{volume}{B139}}, \bibinfo{pages}{170} (\bibinfo{year}{1984}).

\bibitem[{\citenamefont{Antoniadis et~al.}(1987)\citenamefont{Antoniadis,
  Ellis, Hagelin, and Nanopoulos}}]{Antoniadis:1987dx}
\bibinfo{author}{\bibfnamefont{I.}~\bibnamefont{Antoniadis}},
  \bibinfo{author}{\bibfnamefont{J.~R.} \bibnamefont{Ellis}},
  \bibinfo{author}{\bibfnamefont{J.~S.} \bibnamefont{Hagelin}},
  \bibnamefont{and} \bibinfo{author}{\bibfnamefont{D.~V.}
  \bibnamefont{Nanopoulos}}, {``}\bibinfo{title}{{Supersymmetric Flipped
  $SU(5)$ Revitalized}},{''} \bibinfo{journal}{Phys. Lett.}
  \textbf{\bibinfo{volume}{B194}}, \bibinfo{pages}{231} (\bibinfo{year}{1987}).

\bibitem[{\citenamefont{Kim and Kyae}(2007)}]{Kim:2006hw}
\bibinfo{author}{\bibfnamefont{J.~E.} \bibnamefont{Kim}} \bibnamefont{and}
  \bibinfo{author}{\bibfnamefont{B.}~\bibnamefont{Kyae}},
  {``}\bibinfo{title}{{Flipped SU(5) from Z(12-I) orbifold with Wilson
  line}},{''} \bibinfo{journal}{Nucl.Phys.} \textbf{\bibinfo{volume}{B770}},
  \bibinfo{pages}{47} (\bibinfo{year}{2007}), \eprint{hep-th/0608086}.

\bibitem[{\citenamefont{Huh et~al.}(2009)\citenamefont{Huh, Kim, and
  Kyae}}]{Huh:2009nh}
\bibinfo{author}{\bibfnamefont{J.-H.} \bibnamefont{Huh}},
  \bibinfo{author}{\bibfnamefont{J.~E.} \bibnamefont{Kim}}, \bibnamefont{and}
  \bibinfo{author}{\bibfnamefont{B.}~\bibnamefont{Kyae}},
  {``}\bibinfo{title}{{SU(5)(flip) x SU(5)-prime from Z(12-I)}},{''}
  \bibinfo{journal}{Phys. Rev.} \textbf{\bibinfo{volume}{D80}},
  \bibinfo{pages}{115012} (\bibinfo{year}{2009}), \eprint{0904.1108}.

\bibitem[{\citenamefont{Jiang et~al.}(2007)\citenamefont{Jiang, Li, and
  Nanopoulos}}]{Jiang:2006hf}
\bibinfo{author}{\bibfnamefont{J.}~\bibnamefont{Jiang}},
  \bibinfo{author}{\bibfnamefont{T.}~\bibnamefont{Li}}, \bibnamefont{and}
  \bibinfo{author}{\bibfnamefont{D.~V.} \bibnamefont{Nanopoulos}},
  {``}\bibinfo{title}{{Testable Flipped $SU(5) \times U(1)_X$ Models}},{''}
  \bibinfo{journal}{Nucl. Phys.} \textbf{\bibinfo{volume}{B772}},
  \bibinfo{pages}{49} (\bibinfo{year}{2007}), \eprint{hep-ph/0610054}.

\bibitem[{\citenamefont{Jiang et~al.}(2009)\citenamefont{Jiang, Li, Nanopoulos,
  and Xie}}]{Jiang:2008yf}
\bibinfo{author}{\bibfnamefont{J.}~\bibnamefont{Jiang}},
  \bibinfo{author}{\bibfnamefont{T.}~\bibnamefont{Li}},
  \bibinfo{author}{\bibfnamefont{D.~V.} \bibnamefont{Nanopoulos}},
  \bibnamefont{and} \bibinfo{author}{\bibfnamefont{D.}~\bibnamefont{Xie}},
  {``}\bibinfo{title}{{F-SU(5)}},{''} \bibinfo{journal}{Phys. Lett.}
  \textbf{\bibinfo{volume}{B677}}, \bibinfo{pages}{322} (\bibinfo{year}{2009}),
  \eprint{0811.2807}.

\bibitem[{\citenamefont{Jiang et~al.}(2010)\citenamefont{Jiang, Li, Nanopoulos,
  and Xie}}]{Jiang:2009za}
\bibinfo{author}{\bibfnamefont{J.}~\bibnamefont{Jiang}},
  \bibinfo{author}{\bibfnamefont{T.}~\bibnamefont{Li}},
  \bibinfo{author}{\bibfnamefont{D.~V.} \bibnamefont{Nanopoulos}},
  \bibnamefont{and} \bibinfo{author}{\bibfnamefont{D.}~\bibnamefont{Xie}},
  {``}\bibinfo{title}{{Flipped $SU(5) \times U(1)_X$ Models from
  F-Theory}},{''} \bibinfo{journal}{Nucl. Phys.}
  \textbf{\bibinfo{volume}{B830}}, \bibinfo{pages}{195} (\bibinfo{year}{2010}),
  \eprint{0905.3394}.

\bibitem[{\citenamefont{Giudice and Masiero}(1988)}]{Giudice:1988yz}
\bibinfo{author}{\bibfnamefont{G.}~\bibnamefont{Giudice}} \bibnamefont{and}
  \bibinfo{author}{\bibfnamefont{A.}~\bibnamefont{Masiero}},
  {``}\bibinfo{title}{{A Natural Solution to the mu Problem in Supergravity
  Theories}},{''} \bibinfo{journal}{Phys. Lett.}
  \textbf{\bibinfo{volume}{B206}}, \bibinfo{pages}{480} (\bibinfo{year}{1988}).

\bibitem[{\citenamefont{ATLAS}(2016)}]{atlas-vectorlike}
\bibinfo{author}{\bibnamefont{ATLAS}}, {``}\bibinfo{title}{Exotics Combined
  Summary Plots},{''} (\bibinfo{year}{2016}),
  \bibinfo{note}{atlas.web.cern.ch/Atlas/GROUPS/PHYSICS/
  CombinedSummaryPlots/EXOTICS/index.html}.

\bibitem[{\citenamefont{Aaltonen}(2013)}]{CDF:2013jga}
\bibinfo{author}{\bibfnamefont{T.~A.} \bibnamefont{Aaltonen}}
  (\bibinfo{collaboration}{Tevatron Electroweak Working Group, CDF, D0}),
  {``}\bibinfo{title}{{Combination of CDF and DO results on the mass of the top
  quark using up to 8.7 $fb^{-1}$ at the Tevatron}},{''}
  (\bibinfo{year}{2013}), \eprint{1305.3929}.

\bibitem[{\citenamefont{Hinshaw et~al.}(2012)}]{Hinshaw:2012aka}
\bibinfo{author}{\bibfnamefont{G.}~\bibnamefont{Hinshaw}} \bibnamefont{et~al.}
  (\bibinfo{collaboration}{WMAP Collaboration}), {``}\bibinfo{title}{{Nine-Year
  Wilkinson Microwave Anisotropy Probe (WMAP) Observations: Cosmological
  Parameter Results}},{''} (\bibinfo{year}{2012}), \eprint{1212.5226}.

\bibitem[{\citenamefont{Ade et~al.}(2015)}]{Planck:2015xua}
\bibinfo{author}{\bibfnamefont{P.}~\bibnamefont{Ade}} \bibnamefont{et~al.}
  (\bibinfo{collaboration}{Planck}), {``}\bibinfo{title}{{Planck 2015 results.
  XIII. Cosmological parameters}},{''} (\bibinfo{year}{2015}),
  \eprint{1502.01589}.

\bibitem[{\citenamefont{Huo et~al.}(2012)\citenamefont{Huo, Li, Nanopoulos, and
  Tong}}]{Huo:2011zt}
\bibinfo{author}{\bibfnamefont{Y.}~\bibnamefont{Huo}},
  \bibinfo{author}{\bibfnamefont{T.}~\bibnamefont{Li}},
  \bibinfo{author}{\bibfnamefont{D.~V.} \bibnamefont{Nanopoulos}},
  \bibnamefont{and} \bibinfo{author}{\bibfnamefont{C.}~\bibnamefont{Tong}},
  {``}\bibinfo{title}{{The Lightest CP-Even Higgs Boson Mass in the Testable
  Flipped $SU(5) \times U(1)_X$ Models from F-Theory}},{''}
  \bibinfo{journal}{Phys.Rev.} \textbf{\bibinfo{volume}{D85}},
  \bibinfo{pages}{116002} (\bibinfo{year}{2012}), \eprint{1109.2329}.

\bibitem[{\citenamefont{HFAG}(2013)}]{HFAG}
\bibinfo{author}{\bibnamefont{HFAG}} (\bibinfo{year}{2013}),
  \bibinfo{note}{www.slac.stanford.edu/xorg/hfag /rare/2013/ radll/OUTPUT/
  TABLES/radll.pdf}.

\bibitem[{\citenamefont{Khachatryan et~al.}(2015)}]{CMS:2014xfa}
\bibinfo{author}{\bibfnamefont{V.}~\bibnamefont{Khachatryan}}
  \bibnamefont{et~al.} (\bibinfo{collaboration}{LHCb, CMS}),
  {``}\bibinfo{title}{{Observation of the rare $B^0_s\to\mu^+\mu^-$ decay from
  the combined analysis of CMS and LHCb data}},{''} \bibinfo{journal}{Nature}
  \textbf{\bibinfo{volume}{522}}, \bibinfo{pages}{68} (\bibinfo{year}{2015}),
  \eprint{1411.4413}.

\bibitem[{\citenamefont{Aoyama et~al.}(2012)\citenamefont{Aoyama, Hayakawa,
  Kinoshita, and Nio}}]{Aoyama:2012wk}
\bibinfo{author}{\bibfnamefont{T.}~\bibnamefont{Aoyama}},
  \bibinfo{author}{\bibfnamefont{M.}~\bibnamefont{Hayakawa}},
  \bibinfo{author}{\bibfnamefont{T.}~\bibnamefont{Kinoshita}},
  \bibnamefont{and} \bibinfo{author}{\bibfnamefont{M.}~\bibnamefont{Nio}},
  {``}\bibinfo{title}{{Complete Tenth-Order QED Contribution to the Muon
  g-2}},{''} \bibinfo{journal}{Phys.Rev.Lett.} \textbf{\bibinfo{volume}{109}},
  \bibinfo{pages}{111808} (\bibinfo{year}{2012}), \eprint{1205.5370}.

\bibitem[{\citenamefont{Akerib et~al.}(2016)}]{Akerib:2016vxi}
\bibinfo{author}{\bibfnamefont{D.~S.} \bibnamefont{Akerib}}
  \bibnamefont{et~al.}, {``}\bibinfo{title}{{Results from a search for dark
  matter in LUX with 332 live days of exposure}},{''} (\bibinfo{year}{2016}),
  \eprint{1608.07648}.

\bibitem[{\citenamefont{Tan et~al.}(2016)}]{Tan:2016zwf}
\bibinfo{author}{\bibfnamefont{A.}~\bibnamefont{Tan}} \bibnamefont{et~al.}
  (\bibinfo{collaboration}{PandaX-II}), {``}\bibinfo{title}{{Dark Matter
  Results from First 98.7-day Data of PandaX-II Experiment}},{''}
  \bibinfo{journal}{Phys. Rev. Lett.} \textbf{\bibinfo{volume}{117}},
  \bibinfo{pages}{121303} (\bibinfo{year}{2016}), \eprint{1607.07400}.

\bibitem[{\citenamefont{Behnke et~al.}(2012)}]{Behnke:2012ys}
\bibinfo{author}{\bibfnamefont{E.}~\bibnamefont{Behnke}} \bibnamefont{et~al.}
  (\bibinfo{collaboration}{COUPP}), {``}\bibinfo{title}{{First Dark Matter
  Search Results from a 4-kg CF$_3$I Bubble Chamber Operated in a Deep
  Underground Site}},{''} \bibinfo{journal}{Phys. Rev.}
  \textbf{\bibinfo{volume}{D86}}, \bibinfo{pages}{052001}
  (\bibinfo{year}{2012}), \bibinfo{note}{[Erratum: Phys.
  Rev.D90,no.7,079902(2014)]}, \eprint{1204.3094}.

\bibitem[{\citenamefont{Aprile et~al.}(2013)}]{Aprile:2013doa}
\bibinfo{author}{\bibfnamefont{E.}~\bibnamefont{Aprile}} \bibnamefont{et~al.}
  (\bibinfo{collaboration}{XENON100}), {``}\bibinfo{title}{{Limits on
  spin-dependent WIMP-nucleon cross sections from 225 live days of XENON100
  data}},{''} \bibinfo{journal}{Phys. Rev. Lett.}
  \textbf{\bibinfo{volume}{111}}, \bibinfo{pages}{021301}
  (\bibinfo{year}{2013}), \eprint{1301.6620}.

\bibitem[{\citenamefont{Belanger et~al.}(2009)\citenamefont{Belanger, Boudjema,
  Pukhov, and Semenov}}]{Belanger:2008sj}
\bibinfo{author}{\bibfnamefont{G.}~\bibnamefont{Belanger}},
  \bibinfo{author}{\bibfnamefont{F.}~\bibnamefont{Boudjema}},
  \bibinfo{author}{\bibfnamefont{A.}~\bibnamefont{Pukhov}}, \bibnamefont{and}
  \bibinfo{author}{\bibfnamefont{A.}~\bibnamefont{Semenov}},
  {``}\bibinfo{title}{{Dark matter direct detection rate in a generic model
  with micrOMEGAs2.1}},{''} \bibinfo{journal}{Comput. Phys. Commun.}
  \textbf{\bibinfo{volume}{180}}, \bibinfo{pages}{747} (\bibinfo{year}{2009}),
  \eprint{0803.2360}.

\bibitem[{\citenamefont{Djouadi et~al.}(2007)\citenamefont{Djouadi, Kneur, and
  Moultaka}}]{Djouadi:2002ze}
\bibinfo{author}{\bibfnamefont{A.}~\bibnamefont{Djouadi}},
  \bibinfo{author}{\bibfnamefont{J.-L.} \bibnamefont{Kneur}}, \bibnamefont{and}
  \bibinfo{author}{\bibfnamefont{G.}~\bibnamefont{Moultaka}},
  {``}\bibinfo{title}{{SuSpect: A Fortran code for the supersymmetric and Higgs
  particle spectrum in the MSSM}},{''} \bibinfo{journal}{Comput. Phys. Commun.}
  \textbf{\bibinfo{volume}{176}}, \bibinfo{pages}{426} (\bibinfo{year}{2007}),
  \eprint{hep-ph/0211331}.

\bibitem[{\citenamefont{Patrignani et~al.}(2016)}]{Olive:2016xmw}
\bibinfo{author}{\bibfnamefont{C.}~\bibnamefont{Patrignani}}
  \bibnamefont{et~al.} (\bibinfo{collaboration}{Particle Data Group}),
  {``}\bibinfo{title}{{Review of Particle Physics}},{''}
  \bibinfo{journal}{Chin. Phys.} \textbf{\bibinfo{volume}{C40}},
  \bibinfo{pages}{100001} (\bibinfo{year}{2016}).

\bibitem[{\citenamefont{Aad et~al.}(2016)}]{Khachatryan:2016vau}
\bibinfo{author}{\bibfnamefont{G.}~\bibnamefont{Aad}} \bibnamefont{et~al.}
  (\bibinfo{collaboration}{ATLAS, CMS}), {``}\bibinfo{title}{{Measurements of
  the Higgs boson production and decay rates and constraints on its couplings
  from a combined ATLAS and CMS analysis of the LHC pp collision data at $
  \sqrt{s}=7 $ and 8 TeV}},{''} \bibinfo{journal}{JHEP}
  \textbf{\bibinfo{volume}{08}}, \bibinfo{pages}{045} (\bibinfo{year}{2016}),
  \eprint{1606.02266}.

\end{thebibliography}

\end{document}